# Estimation of Rain Attenuation at EHF bands for Earth-to-Satellite Links in Bangladesh

Md. Sakir Hossain
Dept. of Electronic and Telecommunications Engineering
International Islamic University Chittagong
Chittagong, Bangladesh
Email: shakir.rajbd@yahoo.com

Md. Atiqul Islam
Dept. of Electrical and Electronic Engineering
International Islamic University Chittagong
Chittagong, Bangladesh
Email: atiq_atrai@yahoo.com

*Abstract*—Due to heavy congestion in lower frequency bands, engineers are looking for new frequency bands to support new services that require higher data rates, which in turn needs broader bandwidths. To meet this requirement, extremely high frequency (EHF), particularly Q (36 to 46 GHz) and V (46 to 56 GHz) bands, is the best viable solution because of its complete availability. The most serious challenge the EHF band poses is the attenuation caused by rain. This paper investigates the effect of the rain on Q and V bands' performances in Bangladeshi climatic conditions. The rain attenuations of the two bands are predicted for the four main regions of Bangladesh using ITU rain attenuation model. The measured rain statistics is used for this prediction. It is observed that the attenuation due to rain in the Q/V band reaches up to 150 dB which is much higher than that of the currently used $K_a$ band. The variability of the rain attenuation is also investigated over different sessions of Bangladesh. The attenuation varies from 40 dB to 170 dB depending on the months. Finally, the amount of rain fade required to compensate the high rain attenuation is also predicted for different elevation angles.

## I. INTRODUCTION

The usage of wireless communication is following a very sharp uptrend. Particularly, the extensive use of multimedia communication makes frequency bands scarce. This makes lower frequency bands very congested, and the need for attempt to use higher frequency bands is imperative. To materialize the global gigabyte wireless connectivity, satellite communication is deemed as one of the most important tool. In satellite communication, the most widely used frequency band is C band, which is already congested heavily. The Ku and Ka bands will become saturated very soon because the use of these bands is increasing rapidly. This motivates engineers to look for higher frequency bands for getting available bandwidth for future satellite communication to cope with the upcoming tremendous demand of new multimedia services. The two bands of EHF band, namely Q and V bands, are being seen as the possible solution to this tremendous scarcity of bandwidth. EHF bands have a number of advantages, such as broader bandwidth, low interference due to high directivity, low transmission power due to high gain antenna, small antennas and so on [1]. Few examples of EHF satellite systems are INTALSAT F1 and F2, Sicral, Alphasat [1]. However, frequency bands above 10 GHz do not appear as solely blessing, rather these are accompanied with severe rain attenuation. The rough estimate of rain attenuation increase is almost the square of the frequency above C band [2]. To design a satellite communication system in Q/V bands, the estimation of rain attenuation in these bands is a prior task for efficient link design.

The rainfall rate is not same all over the world, instead it varies significantly. The amount of rain attenuation depends on the rainfall rate. So the rain attenuation is not fixed across the world. The rain attenuations in different frequency bands are different. The effects of weather related issues on microwave link in the Bangladeshi climatic condition are investigated in [3]-[7]. The rain attenuation for Bangladeshi climate in C, Ku and Ka bands has been predicted in [3], [6], [7]. Considering the uptrend of the congestion in the lower frequency bands, it is necessary to perform a feasibility study of using higher frequency bands. In this paper, the rain attenuations in the Q and V bands are predicted for Bangladeshi climate to facilitate satellite link design in Bangladesh. It is observed that the attenuation in the Q and V bands are 50 dB and 100 dB higher compared to that in Ka bands for 0.001% time. This requires a very large fade margin to keep the received signal in the EHF band acceptable. The required fade margins are around 150 dB and 110 dB respectively in the Q and V bands, respectively, for 0.01% time of the year. Sylhet region requires the highest fade margin, which is about 20 dB higher compared to the northern region Rajshahi, the least affected region. In addition, the attenuation increases four times in rainy season compared to winter season.

## II. RAIN RATE DISTRIBUTION IN BANGLADESH

Bangladesh has subtropical monsoon climate characterized by wide seasonal variations in rainfall. Heavy rainfall is a feature of Bangladesh. There is a significant variation in the annual rainfall across Bangladesh; this is highest in Sylhet due to its location in the south of the foothills of the Himalayas, while the minimum is experienced in relatively dry western region of Rajshahi. Including the rain attenuation in the mentioned two regions, the attenuation for Dhaka and Chittagong, two most industrial regions where most of the earth stations are located, are predicted here. The outage of a satellite system depends considerably on the signal attenuation due to the rain, which in turn depends on instantaneous rain rate. But the rain rate is not constant over the year. So cumulative distribution (CD) of the rain across the year is used to predict rain attenuation. Here the annual rainfall statistics for forty years (from 1968 to 2008) of four different regions are collected from Bangladesh agriculture research council (BARC) [8], and the average annual rainfall is computed from the data. Then, the CD of rain rate is found using Rice-Holmberg Rain Model (R-H rain model) [9]. The parameters needed for R-H model include Thunderstorm Ratio β (amount of convective rain compared to the total rainfall) and the average annual rainfall depth M. The values of these parameters are shown in Table I, where β is taken from the map in [9].

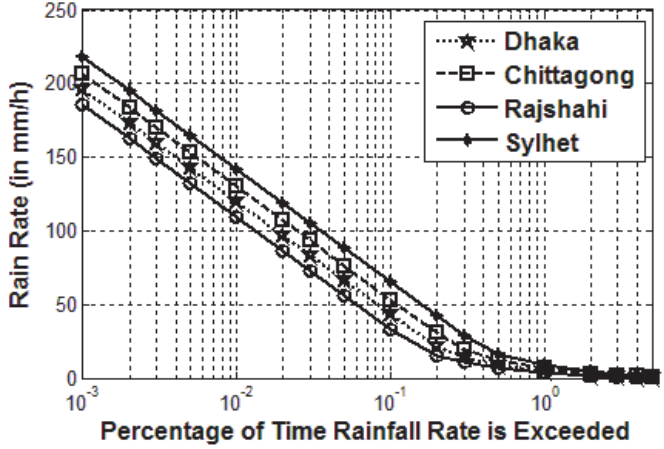

Fig. 1 Cumulative rain distribution of Bangladesh by R-H rain model

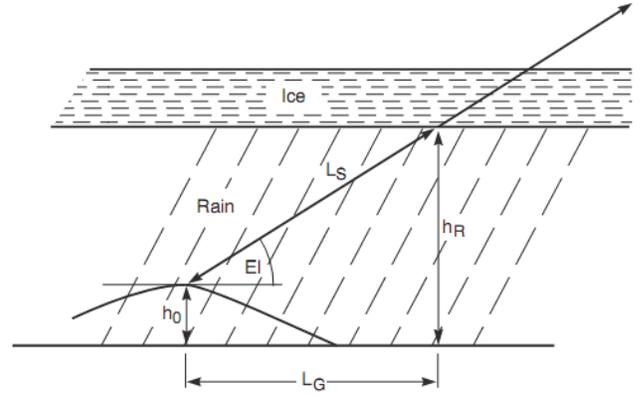

Fig. 2 Schematic presentation of an Earth-space path giving the parameters to be input into the attenuation prediction process [17].

TABLE I. PARAMETERS OF R-H RAIN MODEL

| Parameters | Dhaka | Chittagong | Rajshahi | Sylhet |
|---|---|---|---|---|
| M (mm) | 2124 | 2887 | 1545 | 4101 |
| β | 0.5 | 0.5 | 0.5 | 0.5 |

This model gives the rain rate CD at 1-minute integration time which is shown in Fig. 1 for the parameters mentioned in Table I. The 0.01% rain rates ($R_{0.01}$) of four regions taken from Fig. 1 are listed in Table II, where the maximum $R_{0.01}$, which is 141.661 mm/h, reported in Sylhet, while the minimum of 109.1496 mm/h in Rajshahi. However, the predicted 0.01% rain rate in Bangladesh is 95 mm/h according to the Recommendation ITU-R P.837-6 [10]. A significant difference is observed between the measured $R_{0.01}$ and ITU predicted $R_{0.01}$.

TABLE II. $R_{0.01}$ IN DIFFERENT REGIONS

| Region | $R_{0.01}$(mm/h) |
|---|---|
| Dhaka | 119.7673 |
| Chittagong | 129.9933 |
| Rajshahi | 109.1496 |
| Sylhet | 141.6991 |
| ITU Map | 95 |

III. RAIN ATTENUATION PREDICTION MODEL

There are a number of rain attenuation prediction models [11]. Of them ITU-R [12] and CCIR [13] are designed for world-wide application for wide range of frequencies, rain climates, and elevation angles. The rest of the models mentioned in [11] are designed to meet the local needs. However, ITU-R rain attenuation prediction model is more widely used than CCIR because of its more accuracy. In this paper, ITU-R model will be used for predicting the attenuation in the EHF band. This model can estimate the attenuation due to rain up to the frequency of 55 GHz. The parameters used in this model are listed below:

$R_{0.01}$ : point rainfall rate for the location for 0.01% of an average year (mm/h)
$h_s$ : height above mean sea level of the earth station (km)
$\Theta$ : elevation angle (degrees)
$\varphi$ : latitude of the earth station (degrees)
$F$ : frequency (GHz)
$R_e$ : effective radius of the earth (8500 km)

The geometry is illustrated in Fig. 2.

The steps of this model are outlined below:
*Step 1:* Determine the rain height, $h_R$, using the following formula [14]:
$$h_R = h_o + 0.36 \; km \qquad (1)$$
where $h_o$ is the 0°C isotherm height or freezing height above the mean sea level (km). The average freezing height for different regions of the world can be found in [14].

*Step 2:* For $\theta \geq 5^o$, compute the slant-path length $L_s$, below the rain height from:
$$L_s = \frac{(h_R - h_s)}{\sin\theta} \; km \qquad (2)$$
For $\theta < 5^o$, the following formula is used:
$$L_s = \frac{2(h_R - h_s)}{\left(\sin^2\theta + \frac{2(h_R - h_s)}{R_s}\right)^{1/2} + \sin\theta} \; km \qquad (3)$$
If $(h_R - h_s) \leq 0$, the predicted rain attenuation for any time percentage is zero and the following steps are not required.

*Steps 3:* Calculate the horizontal projection, $L_G$, of the slant-path length from:
$$L_G = L_s \cos\theta \; km \qquad (4)$$

*Step 4:* Obtain the rainfall rate, $R_{0.01}$, exceeded for 0.01% of an average year (with an integration time of 1 minute). If this long-term statistic cannot be obtained from local data sources, an estimate can be obtained from the maps of rainfall rate given in [10]. If $R_{0.01} = 0$, the predicted rain attenuation is zero for any time percentage and the following steps are not required.

*Step 5:* Obtain the specific attenuation, $\gamma_R$, using the frequency-dependent coefficients given in [11] and the rainfall rate, $R_{0.01}$, determined in Step 4, by using:

$$\gamma_R = k(R_{0.01})^\alpha \ dB/km \quad (5)$$

*Step 6:* Calculate the horizontal reduction factor, $r_{0.01}$, for 0.01% of the time:

$$r_{0.01} = \frac{1}{1+0.78\sqrt{\frac{L_G\gamma_R}{f}}-0.38(1-e^{2L_G})} \quad (6)$$

Step 7: Calculate the vertical adjustment factor, $v_{0.01}$, for 0.01% of the time:

$$\xi = tan^{-1}\left(\frac{h_R - h_s}{L_G r_{0.01}}\right) \ degrees$$

For $\xi > \theta$, $\quad L_R = \frac{L_G r_{0.01}}{cos\theta} \ km$

else, $\quad L_R = \frac{h_R - h_s}{sin\theta} \ km$

If $|\varphi| < 36^o$, $\quad \chi = 36 - |\varphi| \ degrees$
else, $\quad \chi = 0 \ degrees$

$$v_{0.01} = \frac{1}{1 + \sqrt{sin\theta}\left(31(1-e^{-(\theta/(1+\chi))})\frac{\sqrt{L_R\gamma_R}}{f^2} - 0.45\right)}$$

*Step 8:* The effective path length: $L_E = L_R v_{0.01} \ km$ (7)

*Step 9:* The predicted attenuation exceeded for 0.01% of an average year is given by

$$A_{0.01} = \gamma_R L_L \ dB \quad (8)$$

*Step 10:* The estimated attenuation to be exceeded for other percentage of an average year, in the range from 0.001% to 5%, is determined from the attenuation to be exceeded for 0.01% of an average year:

If $p \geq 1\%$ or $|\varphi| \geq 36^o$, $\beta = 0$

If $p < 1\%$ and $|\varphi| < 36^o$ and $\theta \geq 25^o$, $\quad \beta = -0.005(|\varphi| - 36)$

Otherwise; $\beta = -0.005(|\varphi| - 36) + 1.8 - 4.25 sin\theta$

$$A_p = A_{0.01}\left(\frac{p}{0.01}\right)^{-(0.655+0.0331\ln(p)-0.045\ln(A_{o.o1})-\beta(1-p)sin\theta)} dB \quad (8)$$

This method provides an estimate of the long-term statistics of the attenuation due to rain.

IV. RAIN ATTENUATION PREDICTION

From the previous section, it is clear that the rain attenuation is a function of a number of factors, including rain rate, elevation angle, frequency, $0^oC$ isotherm height, height of the earth station from the mean sea level, and so on. If any of these parameters gets changed, the corresponding rain attenuation will also change. This is why determining the values of these parameters is a prerequisite for rain attenuation prediction. Considering the longitude (78.5 (-180o to -180o)) of the satellite THAICOM-5, the parameters for four different regions of Bangladesh are shown in Table III, where the values of the first three parameters are taken from [18], and the corresponding antenna elevation angle for THAICOM-5 satellite is computed using elevation angle calculator [19].

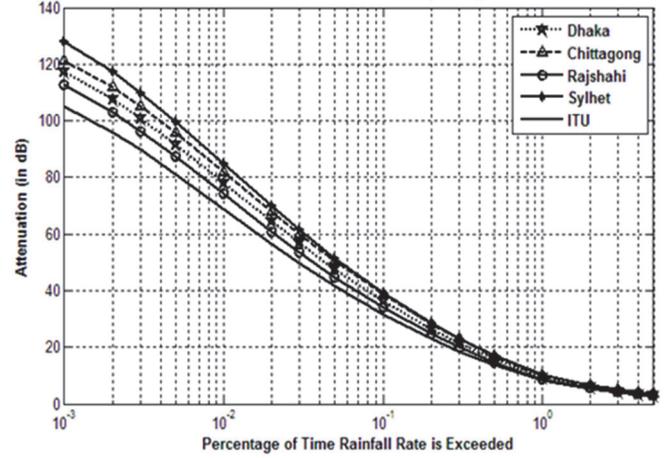

Figure-3: Rain Attenuation at $K_a$ Band (30 GHz)

The 0°C isotherm height varies significantly from 1 km to 13 km in Bangladesh, However most of the precipitation occurs with freezing height between 3 to 5 km [6]. However, according to ITU-R P.839-3 [14], the average freezing height for the Bangladesh is 4.5 km. Since the freezing height varies considerably with the season of the year, the value of this parameter is taken from [14]. The regression coefficients, namely k and α, for Ka, Q, and V bands are taken from [15] for simulation. The frequencies of Q and V bands used in simulation are taken from the US Federal Communication Communications (FCC)'s proposed frequency bands: 37.5-40.5 (uplink) and 47.2-50.2 (downlink) for geostationary satellite, and 37.5-38.5 (uplink) and 48.2-49.2 (downlink) for non-geostationary satellite [16]. The 40 GHz (uplink frequency) of Q band and 50 GHz (downlink frequency) of V band are used for simulation.

TABLE III. SIMULATION PARAMETERS

| Parameters | Dhaka | Chittagong | Rajshahi | Sylhet |
|---|---|---|---|---|
| Longitude (degree) | 90.24 | 91.5 | 88.36 | 91.52 |
| Latitude (degree), $\varphi$ | 23.42 | 22.19 | 24.22 | 24.53 |
| height above mean sea level of the earth station (km), $h_s$ | $4\times10^{-3}$ | $7\times10^{-3}$ | $31\times10^{-3}$ | $9\times10^{-3}$ |
| Antenna Elevation Angle (degree), $\theta$ | 59.58 | 60.15 | 59.59 | 57.82 |

Fig. 3 shows the rain attenuation exceeded at different percentage of time of the year for $K_a$. The maximum rain attenuation of 85 dB is observed for Sylhet, and the minimum of 74 dB for Rajshahi exceeded for 0.01% time of the year. However, the rain attenuation predicted from ITU rain rate is 69 dB exceeded for 0.01% time of the year, which is far lower than the lowest rain attenuation observed in Rajshahi among these four regions from measured rain rate. The rain attenuation at Q and V bands are shown in Fig. 4 and Fig. 5. Since the rain rate is maximum at Sylhet, the maximum rain attenuations are observed here for both Q and V bands, which are 124 dB and 156 dB, respectively. On the other hand, the minimum rain

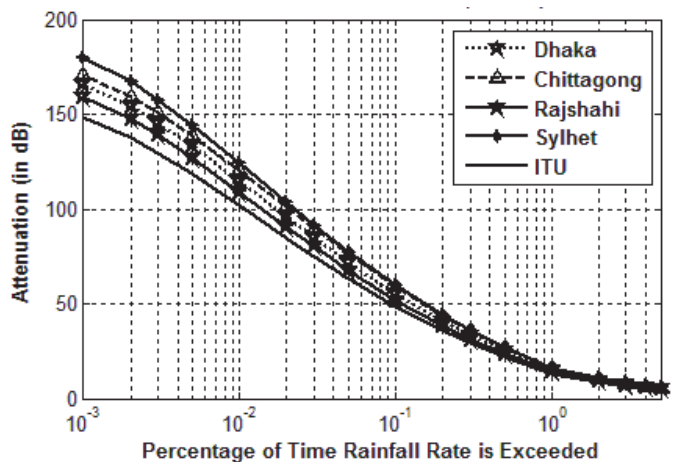
Fig. 4 Rain attenuation at Q-band (40 GHz)

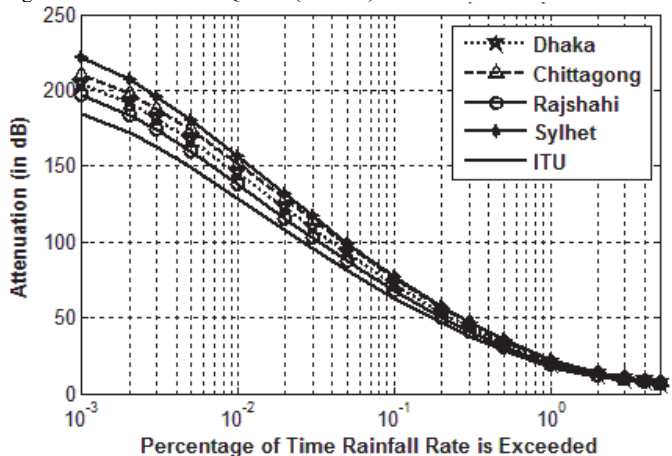
Fig. 5 Rain attenuation at V-band (50 GHz)

attenuations of 109 dB and 138 dB are found in Q and V bands for Rajshahi where the rain rate is minimum. But the rain attenuation predicted from the ITU rain rate in Q and V bands are 102 dB and 129 dB, respectively. There is a significant difference in rain attenuation across Bangladesh. While the maximum attenuation is observed at Sylhet, the minimum at Rajshahi. The difference in rain attenuation exceeded for 0.01% time of the year between these two regions is about 12 % for both bands. A significant difference between the rain attenuation predicted from the measured rain rate and ITU predicted rain rate is found. The rain attenuation predicted using ITU rain rate data is 6% lower than the least rain attenuation in Bangladesh found in Rajshahi and about 18% lower than the highest rain attenuation in Bangladesh observed in Sylhet. The rain attenuation exceeded for 0.01% time of the year in V band is observed in Q band for 0.03% time of the year, while the attenuation exceeded for 0.01% time of the year in Q band is same as the 0.001% time of the year for $K_a$ band.

The period of time at which a network remains in outage stage is an important parameter for determining the quality of the network. This outage can occur due to the system failure or degradation of signal quality below a certain threshold. For the latter one, the signal degradation can occur because of the rain attenuation. To cope with this, a fade margin of a certain

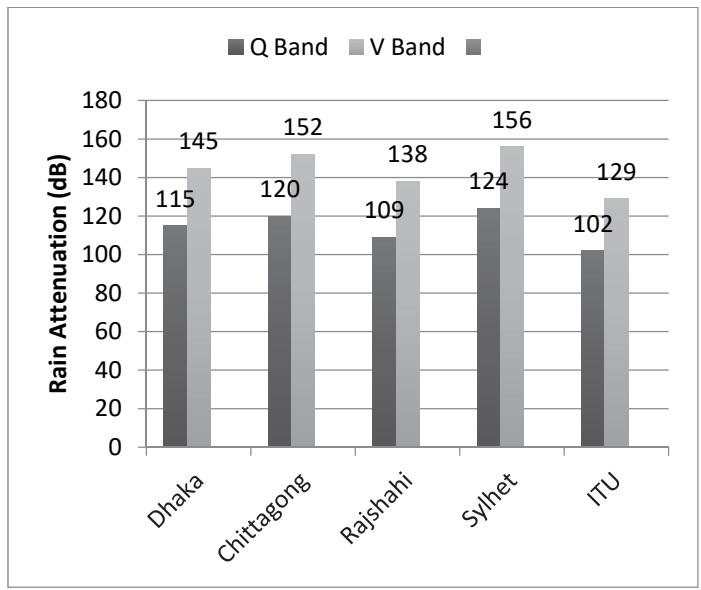
Fig. 6 Average rain fade margin for 99.99% availability.

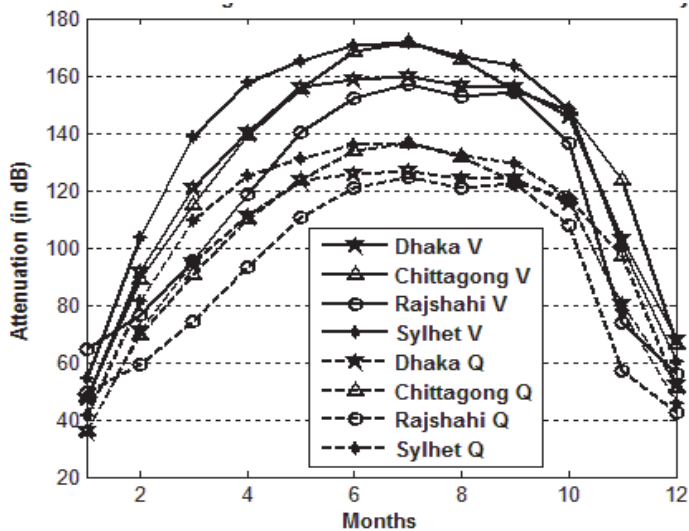
Fig. 7 Rain fade margin for 99.99% availability in different months for Q and V bands.

amount is provided. Fig. 6 provides the average fade margin required for 99.99% availability (0.01% outage) of satellite network for Q and V bands. According to the column chart, the maximum of 156 dB and 124 dB fade margin are required in Sylhet region for V and Q bands, respectively, to ensure 99.99% availability, while the minimum of 138 dB and 109 dB in Rajshahi. However, there is a significant difference in fade margin predicted from ITU predicted rain data and the measured rain data.

The rain attenuation varies according to the amount of precipitation. There is a huge variation of precipitation across the year. The maximum rain occurs in July, but the minimum in January. For this reason, the fade margin required to maintain the certain amount of signal quality varies over months. Fig. 7 shows how the required fade margin varies over months for 99.99% availability in the four regions. The month of July requires the highest fade margin of about 170 dB for V band in Sylhet and Chittagong and the same for Q band is 145 dB in the same regions, though the minimum fade margin, which is 40-60 dB, is required in January. The rain fade margin remains more than 140 dB and 100 dB for 8 months from March to October for V band and Q band, respectively.

V. CONCLUSION

Attenuation due to rain is a severe bar in designing satellite link in frequency more than 10 GHz. There is a large variation in precipitation across Bangladesh and the rain fall rate in monsoon season is 4 to 8 times higher compared to that in dry season. For this reason, the rain attenuation is not same in all regions of Bangladesh, rather it varies from season to season. The rain attenuations in Q and V bands are much higher (about 32% in Q band and 46% in V band) than the existing Ka band. The difference in the attenuation across the country is about 12%, where the maximum is observed at Sylhet and the least value at Rajshahi region. It is also found that the ITU rain rate prediction model cannot predict the rain rate properly for a subtropical country like Bangladesh, because it is found that its predicted rain rate is 6% lower than the forty years average rain rate of the driest region of Bangladesh, and the corresponding effect on the rain attenuation predicted from this predicted rain rate cannot provide the accurate value. To ensure 99.99% availability of network, the required highest average fade margin of 156 dB and 124 dB are reported in Sylhet and the minimum values of 138 dB and 109 dB in Rajshahi for V and Q bands, respectively. The required fade margin varies considerably across the year. The required margin for 99.99% is more than 140 dB for V band and 100 dB for Q band for about six months. Since providing such a huge link margin is somewhat economically impractical with the current technology, the availability will vary with the season in a year. During dry season, higher availability can be enjoyed, but in monsoon this will go down. However, placing satellite in a lower longitude increases the earth station elevation angle, which reduces the amount of fade margin required. In addition, a satellite can be launched in which the total number of transponder will be grouped into two categories: one operating in lower frequency band and the other in Q/V band. During the monsoon season, the lower frequency band will be used, and the EHF band will be in operating mode in the dry season. Transmitting higher power during rain and reducing that in other time can also reduce the effect of rain in the quality of satellite services.